\documentclass[11pt,a4paper,twoside,groupcitations]{article}
\usepackage[T1]{fontenc}
\usepackage[ansinew]{inputenc}
\usepackage[english]{babel}
\usepackage{amsfonts}
\usepackage{amsmath}
\usepackage{bm}
\usepackage{array}
\usepackage{amsthm}
\usepackage{amssymb}
\usepackage{graphicx}
\usepackage{subfigure}
\usepackage{braket}
\usepackage{eucal}
\usepackage{verbatim}
\usepackage[table]{xcolor}
\usepackage{caption}
\usepackage{cite}
\usepackage{textcomp}
\usepackage{multicol}
\usepackage{tikz}
\usetikzlibrary{positioning,arrows}
\usetikzlibrary{decorations.pathmorphing}
\usetikzlibrary{decorations.markings}
\usetikzlibrary{calc,decorations.markings}
\usetikzlibrary{arrows,shapes}
\usetikzlibrary{matrix,arrows}
\usepackage{pgfplots}
\usepackage{xparse}
\definecolor{jade}{HTML}{00A86B}
\newcommand{\be}{\begin{eqnarray}}
\newcommand{\ee}{\end{eqnarray}}
\renewcommand{\d}{\mbox{${\rm d}$}} %d differenziale non corsivo in math mode
\newcommand{\lp}{\ell_{\rm p}}
\newcommand{\mpl}{m_{\rm p}}
\newcommand{\gn}{G_{\rm N}}

\def\beq{\begin{equation}}
\def\eeq{\end{equation}}
\def\@fmsl@sh#1#2#3{\m@th\ooalign{$\hfil#1\mkern#2/\hfil$\crcr$#1#3$}}
 \def\eq#1\en{\begin{equation}#1\end{equation}}
\def\s[#1,#2]{[#1\stackrel{\star}{,}#2]}
\def\sx[#1,#2]{[#1\stackrel{\star_{x}}{,}#2]}

\def\beq{\begin{equation}}
\def\eeq{\end{equation}}
\DeclareMathOperator{\Tr}{Tr}
\title{\bf Bootstrapped Newtonian quantum gravity}
\author{Roberto~Casadio$^{ab}$\thanks{E-mail: casadio@bo.infn.it}
$\ $
and
Iber\^e~Kuntz$^{ab}$\thanks{E-mail: kuntz@bo.infn.it}
\\
\\
$^a${\em Dipartimento di Fisica e Astronomia, Universit\`a di Bologna}
\\
{\em via Irnerio~46, 40126 Bologna, Italy}
\\
\\
$^b${\em I.N.F.N., Sezione di Bologna, I.S.~FLAG}
\\
{\em viale B.~Pichat~6/2, 40127 Bologna, Italy}
}
\begin{document}
\maketitle
\begin{abstract}
We compute quantum corrections for the gravitational potential obtained by including a derivative self-coupling
in its classical dynamics as a toy model for analysing quantum gravity in the strong field regime.
In particular, we focus on quantum corrections to the classical solutions in the vacuum outside localised 
matter sources.
\end{abstract}  
%
%
%\pacs{}
%
%
%%%%%%%%%%%%%%%%%%%%%%%%%%%%%%%%%%%%%%%%%%%%%%%%%%%%%%%%%%%%%%%%
%%%
%%%                     INTRODUCTION
%%%
%%%%%%%%%%%%%%%%%%%%%%%%%%%%%%%%%%%%%%%%%%%%%%%%%%%%%%%%%%%%%%%%
%
\newpage
\section{Introduction}
\label{sec:intro}
\setcounter{equation}{0}
In quantum field theory, fundamental forces are associated with the exchange of (virtual) quanta
of the interaction fields among matter fields and static potentials only emerge as approximate descriptions for particular configurations.
For instance, the linear interaction among fermions in Quantum Electro-Dynamics (QED) is carried by the two polarisations of the
massless vector field whose quanta are the photons. 
The Coulomb potential describing the force acting on a (strictly speaking) static charge then emerges in the non-relativistic
limit of the tree-level transition amplitude for the scattering between two charged fermions via the exchange of these (virtual)
photons~\cite{Itzykson:1980rh}.
For extended sources involving many charged particles, the calculation of amplitudes becomes immediately very 
cumbersome already at the tree level, and things only get worse when nonlinearities stemming from quantum (loop) corrections
are included.
\par
The linear Newtonian interaction likewise emerges from a suitable limit for the exchange of spin~2 (virtual) gravitons between two massive
particles in the weak field regime.
However, according to General Relativity (GR), nonlinearities should already be present at the classical level, which makes explicit
quantum calculations for configurations in the strong field regime very difficult, if possible at all.
As an effective description of the gravitational force, say applied on a test particle by an extended matter source,
one can instead consider the static potential as the mean field generated by that extended source and quantise it canonically. 
This approach applied to QED leads to the description of a static electric field in terms of the coherent state of (virtual)
longitudinal photons~\cite{Mueck:2013wba}.
It then appears straightforward that one can quantise the Newtonian potential, which solves the classical Poisson equation,
in terms of (virtual) scalar gravitons, in a similar fashion.
However, for the purpose of studying the strong field regime of gravity, it is more interesting to try and include some
nonlinearities~\cite{Casadio:2016zpl} in the classical equation for the gravitational potential (the bootstrapped Newtonian
gravity introduced in Refs.~\cite{Casadio:2018qeh,Casadio:2019cux}) and then express the resulting solutions in terms of modified
coherent states~\cite{Casadio:2017cdv,Casadio:2020ueb}.
\par
In this work, we will instead quantise the bootstrapped potential as a scalar field in order to determine one-loop quantum corrections
to its effective field equation.
For this purpose, we will first perform a field redefinition to dispose of the derivative self-interaction and then
apply heat kernel techniques in order to compute the quantum effective
action~\cite{Vilkovisky:1984st,Ellicott:1987ir,Vassilevich:2003xt,Barvinsky:1985an,Barvinsky:1987uw,Barvinsky:1990up}.  
Hopefully, this simplified approach to nonlinearities will help to gain some insight about the quantum dynamics of gravity
in the strong field regime generated by the presence of matter sources.
\par
This paper is organised as follows:
in Section~\ref{sec:redef}, we show how we can deal with derivative interactions non-perturbatively by means of a field
redefinition that is able to transform the action into a canonical form;
in Section~\ref{sec:covaction}, we review the unique effective action first introduced in Ref.~\cite{Vilkovisky:1984st}
and which allows us to extend the findings of Section~\ref{sec:redef} to the quantum level by requiring the quantum
action to be covariant off-shell;
Section~\ref{sec:boot} is devoted to the calculation of the quantum action for bootstrapped gravity.
We show that the quantum equation of motion is described by a non-local equation in the infrared regime,
which is then solved for an idealized point-like source and for a pair of point-like sources;
we finally draw some conclusions in Section.~\ref{sec:conc}
\section{Field redefinition}
\label{sec:redef}
\setcounter{equation}{0}
We start by recalling that the action for the bootstrapped Newtonian potential $V=V(r)$ for spherically 
symmetric systems is given by~\cite{Casadio:2018qeh}~\footnote{We shall use units with $c=1$,
the Newton constant $\gn=\lp/\mpl$ and $\hbar=\lp\,\mpl$.}
\be
S[V]
=
-4\,\pi\int
r^2\,\d r
\left[
\frac{(V')^2}{8\,\pi\,\gn}\left(1-4\,q_\phi\,V\right)
+q_{\rm B}\,\rho\,V\left(1-2\,q_\phi\,V\right)
\right]
\ , 
\ee
where $q_\phi$ is a (positive) coupling that controls the potential self-interaction and $q_{\rm B}$ is introduced to
keep track of the coupling with the matter source of density $\rho$.~\footnote{In the present work we neglect the
pressure term analysed in Ref.~\cite{Casadio:2019cux}, and just note that it could simply be absorbed into the
definition of the matter density.} 
We here want to show that the derivative interaction can be transformed into a canonical kinetic term under a
field redefinition, which is only possible because the configuration space turns out to have vanishing curvature.
This will allow us to quantize the theory non-perturbatively in $q_\phi$, although we will still need to invoke
perturbation theory in the coupling constant $q_{\rm B}$ to deal with matter interactions.
\par
We first rescale the (dimensionless) gravitational potential~\cite{Casadio:2017cdv}
\be
\phi
=
\frac{V}{\sqrt{\gn}}
=
\sqrt{\frac{\mpl}{\lp}}\,V
\ ,
\label{phiV}
\ee
the matter density
\be
J_{\rm B}
=
4\,\pi\,\sqrt{\gn}\,\rho
=
4\,\pi\,\sqrt{\frac{\lp}{\mpl}}\,\rho
\ ,
\label{rhoJ}
\ee
and promote the new scalar field $\phi=\phi(x^\mu)$ as well as $J_{\rm B}=J_{\rm B}(x^\mu)$ to depend on all spacetime
coordinates $x^\mu=(t,\vec x)$. 
The generalised bootstrapped action then reads
\beq
S[\phi,J_B] = S_0[\phi] + S_\text{int}[\phi,J_B]
\ ,
\label{eq:frame1}
\eeq
where 
\be
\label{eq:s0}
S_0[\phi] = \int\mathrm{d}^4x \left[-\frac12\partial_\mu\phi\,\partial^\mu\phi + \alpha\,(\partial_\mu\phi)^2\,\phi\right]
\ee
is the kinetic part containing a derivative self-interaction, and 
\be
S_\text{int}[\phi,J_B] 
=
\int\mathrm{d}^4x\, \xi(\phi)\,J_{\rm B}
\ .
\ee
In order to avoid heavy notation, we also introduced
\be
\alpha = 2\,q_\phi\,\sqrt{\frac{\lp}{\mpl}}
\label{alpha}
\ee
and
\be
\xi(\phi) = -q_{\rm B}\, \phi\left(1-\alpha\,\phi\right)
\ ,
\ee
which represents a non-linear coupling to the source $J_{\rm B}$.
\par
The above action contains derivative interactions, which means that the action for the free field is not recovered 
by simply setting $q_{\rm B}=q_\phi=0$.
But one can perform a field redefinition and try to put it in canonical form by diagonalising the whole kinetic Lagrangian
density
\be
\mathcal L_0 = \left(-\frac12 + \alpha\,\phi\right)\partial_\mu\phi\,\partial^\mu\phi
\ .
\label{eq:lag}
\ee
As can be seen from Eq.~\eqref{eq:lag}, for $\phi>{1}/{2\,\alpha}$ the kinetic term changes sign and $\phi$ becomes a ghost.
For this reason, we will focus on the branch $\phi<{1}/{2\,\alpha}$.~\footnote{Typical classical solutions are expected
to have $\phi<0$, so that this condition is trivially satisfied.}
The dependence on $\phi$ inside the brackets that multiply $(\partial_\mu\phi)^2$ also indicates that the metric in field space
is not trivially flat, which can either mean that the field space is curved or that the field space is flat but the chosen coordinate
is curvilinear.
If the field space is flat, then there exists a field transformation which diagonalises the kinetic term.
Reciprocally, should there be a frame where the kinetic term is diagonal, then the field space must be flat.
We will show below that this is indeed the case for the Lagrangian density~\eqref{eq:lag}.
\par
To put the kinetic term in the canonical form, we need a field redefinition $\varphi=\psi(\phi)$ such that
\be
\partial_\mu\varphi = \sqrt{1-2\, \alpha\,\phi}\,\partial_\mu\phi
\ ,
\ee
which is real and non-singular for
\beq
\phi
<
\frac{1}{2\,\alpha}
\ ,
\label{noghost}
\eeq
and is solved by
\beq
\varphi 
=
C
-\frac{1}{3\,\alpha}
\left(
1-2\,\alpha\,\phi
\right)^{3/2}
\ ,
\eeq
where $C$ is an arbitrary integration constant.
Upon requiring that the transformation reduces to the identity for $\alpha\to 0$, we obtain
\beq
\varphi
=
\psi(\phi) 
=
\frac{1}{3\,\alpha}
\left[
1
-
\left(
1-2\,\alpha\,\phi
\right)^{3/2}
\right]
\ .
\label{vphiPhi}
\eeq
Upon inverting the above relation, we get
\beq
\phi 
=
\psi^{-1}(\varphi)
=
\frac{1}{2\,\alpha}
\left[
1
-
\left(
1-3\,\alpha\,\varphi
\right)^{2/3}
\right]
\ ,
\label{eq:redef}
\eeq
which is precisely the relation between the exact ``vacuum'' solution
\beq
V_{\rm c}
=
\frac{1}{4\,q_\phi}
\left[
1
-
\left(
1-6\,q_\phi\,V_{\rm N}
\right)^{2/3}
\right]
\label{Vc}
\eeq
and the Newtonian potential
\beq
V_{\rm N}
=
-\frac{G_{\rm N}\,M}{r}
\ ,
\label{Vn}
\eeq
where $M$ is the source mass.~\footnote{We shall see later on that the mass $M$ however differs from the Newtonian
expectation~\cite{Casadio:2018qeh}.}
This is consistent with the fact that the Newtonian potential has canonical kinetic term.
In terms of $\varphi$, the complete Lagrangian density then reads
\beq
\mathcal L
=
-\frac12\,\partial_\mu\varphi\,\partial^\mu\varphi
+ \tilde \xi(\varphi)\,J_{\rm B}
\ ,
\label{eq:diag}
\eeq
where the non-linear coupling $\tilde\xi(\varphi)=\xi(\psi^{-1}(\varphi))$ is given by
\be
\tilde\xi(\varphi)
=
-\frac{q_{\rm B}}{4\,\alpha}
\left[1
-(1-3\,\alpha\,\varphi)^{4/3}
\right]
\ .
\ee
Since the interaction terms do not contain any derivatives, the Lagrangian density~\eqref{eq:diag} for $\varphi$ can be
quantised in the standard way by defining the asymptotic states for the free field $\varphi_0$.
\section{Covariant quantum action}
\label{sec:covaction}
\setcounter{equation}{0}
The quantisation of the theory will be performed using the one particle irreducible (1PI, or quantum) action.
The calculation of the 1PI action in our case is a little subtler than the standard calculations in quantum field theory
because of the curvilinear coordinates in the original frame~\eqref{eq:frame1}.
In particular, the naive definition of the 1PI action is frame dependent off-shell and can only be used if one is interested
in the components of the S-matrix, which are calculated on-shell.
On the other hand, we are interested in the dynamics of the off-shell mean field in a given state,
whose quantum action would produce non-covariant results.
Fortunately, a covariant formulation of the 1PI action exists as it was introduced in Ref.~\cite{Vilkovisky:1984st}
(see also~\cite{Ellicott:1987ir}).
In the following, we briefly review such formulation in one-dimensional field space, which is enough for our purposes
since the action~\eqref{eq:frame1} contains only one degree of freedom.
\par
The covariant partition function is given by
\beq
Z[J]
=
\int\mathcal{D}\phi\, e^{i\,\left\{ S[\phi] + \left[v[\bar\phi] - \sigma(\bar\phi,\phi)\right]J \right\}}
\ ,
\label{eq:covpf}
\eeq
where $v[\bar\phi]$ is an arbitrary covector field satisfying $\nabla_{\bar\phi}v[\bar\phi] = 1$ with $\nabla_{\bar\phi}$ being the covariant derivative associated with the Levi-Civita connection $\gamma(\bar\phi)$ of the field space,
the displacement vector is defined by $\sigma(\bar\phi,\phi)=\partial^{\bar\phi}\tilde\sigma(\bar\phi,\phi)$
and $\tilde\sigma(\bar\phi,\phi)$ is Synge's world function, which is numerically equal to one-half of the
square of the geodesic distance between $\phi$ and $\bar\phi$.
Because $v[\bar\phi]$ transforms as a covector and $\sigma(\bar\phi,\phi)$ transforms as a covector
with respect to its first argument and as a scalar with respect to its second argument, the partition function
$Z[J]$ is completely covariant under redefinitions of the background and of the quantum fields.
The covariant relation between the background field and the mean field is now given by
\beq
\left<\sigma(\bar\phi,\phi)\right>
=0
\ ,
\eeq
instead of the naive relation $\bar\phi = \left<\phi\right>$.
The displacement vector can be expanded as
\beq
-\sigma(\bar\phi,\phi)
=
\phi-\bar\phi + \frac12\,\gamma(\bar\phi)\left(\phi-\bar\phi\right)^2
+
\mathcal O(\phi-\bar\phi)^3
\ ,
\eeq
where $\gamma(\bar\phi)$ is the one-dimensional Levi-Civita connection in field space,
thus one obviously retrieve the naive definition $\bar\phi = \left<\phi\right>$ for a vanishing connection.
The covariant 1PI action is defined as
\be
\label{eq:legendre}
\Gamma[\bar\phi] = W[J] - v[\bar\phi]\,J
\ ,
\quad
{\rm with}
\quad
\frac{\delta W[J]}{\delta J} = v[\bar\phi]
\ ,
\ee
where $W[J] = -i\,\log Z[J]$ is the generating functional of connected diagrams.
From Eqs.~\eqref{eq:covpf} and \eqref{eq:legendre}, one obtains
\be
e^{i\,\Gamma[\bar\phi]}
=
\int\mathcal{D}\phi \,e^{i\left\{ S[\phi] + \sigma(\bar\phi,\phi)\,\nabla_{\bar\phi}\Gamma[\bar\phi] \right\}}
\ .
\label{eq:uniqueact}
\ee
To perform the integral in Eq.~\eqref{eq:uniqueact}, it is best to make the change of variables
\be
\phi\to \sigma(\bar\phi,\phi)
\ .
\ee
By expanding the classical action as a covariant Taylor series
\be
S[\phi]
=
\sum^\infty_{n=0}\frac{(-1)^n}{n!}\left\{\nabla_{\sigma}S[\bar\phi]\,\sigma(\bar\phi,\phi)\right\}^n
\ee
and the quantum action as a loop expansion
\be
\Gamma[\bar\phi]
=
S[\bar\phi] + \hbar\, \Gamma^{(1)}[\bar\phi] + \hbar^2\, \Gamma^{(2)}[\bar\phi] + \ldots
\ ,
\ee
one can calculate the quantum action order by order in $\hbar$.
At one-loop order, we find
\be
\Gamma[\bar\phi]
=
S[\bar\phi]
+\frac{i\,\hbar}{2}\,\Tr\log\left(\nabla^2 S[\bar\phi]\right)
\ ,
\ee
where $\Tr$ denotes the functional trace.
The only difference with respect to the standard quantum action is the appearance of the functional covariant derivative
instead of the standard functional derivative, thus making $\Gamma[\bar\phi]$ a covariant object under field redefinitions.
\section{Quantum action for bootstrapped gravity}
\label{sec:boot}
\setcounter{equation}{0}
The result in the previous section could be applied to the Lagrangian density~\eqref{eq:diag} by interpreting $J=J_{\rm B}$
as the auxiliary source.
An alternative procedure would rather be to introduce a different auxiliary source $J$,~\footnote{The source $J_{\rm B}$
is meant to generate a non-trivial background field and cannot be made to vanish arbitrarily.}
in which case the partition function would be given by
\be
Z[J,J_B]
=
\int\mathcal{D}\phi\, e^{i\left\{ S[\phi,J_B] + \left[v[\bar\phi] - \sigma(\bar\phi,\phi)\right]J \right\}}
\ .
\ee
In this way, $J$ can be simply interpreted like a Lagrange multiplier, which will be set to zero in the end as usual,
and needs not necessarily be a physical source.
The 1PI action results in
\begin{align}
e^{i\,\Gamma[\bar\phi,J_{\rm B}]}
&=
\int\mathcal{D}\phi\, e^{i\left\{ S[\phi,J_{\rm B}]
+ \sigma(\bar\phi,\phi)\nabla_{\bar\phi}\Gamma[\bar\phi,J_{\rm B}] \right\}}
\\
&=
\int\mathcal{D}\varphi \,e^{i\left\{ \tilde S[\varphi,J_{\rm B}]
+ \sigma(\psi^{-1}(\bar\varphi),\psi^{-1}(\varphi))\,\nabla_{\bar\varphi}\tilde\Gamma[\bar\varphi,J_{\rm B}] \right\}}
\ ,
\end{align}
where in the second line we used the redefinition~\eqref{eq:redef} and denoted
\be
\tilde S[\varphi,J_{\rm B}]
=
S[\psi^{-1}(\varphi),J_{\rm B}]
=
\int\mathrm{d}^4x\left[ -\frac12\,\partial_\mu\varphi\,\partial^\mu\varphi
+ \tilde \xi(\varphi)\,J_{\rm B}\right]
\ ,
\ee
and $\tilde \Gamma[\varphi,J_{\rm B}] = \Gamma[\psi^{-1}(\varphi),J_{\rm B}]$.
As it was already pointed out in Section~\ref{sec:redef}, in the new frame the field coordinates are Cartesian.
The covariant functional derivative thus reduces to the flat form $\nabla_\varphi = \frac{\delta}{\delta\varphi}$,
and the displacement vector reduces to the coordinate difference in field space
\be
\sigma(\psi^{-1}(\bar\varphi),\psi^{-1}(\varphi))
=
\psi^{-1}(\bar\varphi) - \psi^{-1}(\varphi)
\ .
\ee
The quantum action to one-loop order then reads
\beq
\tilde\Gamma[\bar\varphi,J_{\rm B}]
=
\tilde S[\bar\varphi,J_{\rm B}]
+ \frac{i\,\hbar}{2}\,\Tr\log\Delta
\ ,
\label{eq:qaction2}
\eeq
with
\begin{align}
\Delta
&=
\Box + \tilde\xi''(\bar\varphi)\,J_{\rm B}
\nonumber
\\
&=
\Box
+ q_{\rm B}\,\alpha\left(1 - 3\,\alpha\,\bar\varphi\right)^{-2/3}
J_{\rm B}
\ ,
\label{eq:delta}
\end{align}
where $\Box = \eta^{\mu\nu}\partial_\mu\partial_\nu$ denotes the D'Alembert operator in flat spacetime.
To calculate the second term in Eq.~\eqref{eq:qaction2}, we use the Schwinger proper time method
to represent the one-loop functional determinant in terms of the heat kernel $K$ of the operator $\Delta$
as
\be
\Tr\log\Delta
=
-\int_0^\infty\d s\, \frac{\Tr K(s)}{s}
\ .
\label{eq:proptime}
\ee
In the presence of a potential term in $\Delta$, such as $P\equiv\tilde\xi''(\bar\varphi)\,J_{\rm B}$ in Eq.~\eqref{eq:delta},
the computation of the exact heat kernel becomes highly non-trivial and it is necessary to rely on approximate methods.
Different approximation techniques with different scopes of applicability have been
developed~\cite{Vassilevich:2003xt,Barvinsky:1985an,Barvinsky:1987uw,Barvinsky:1990up},
the most popular one perhaps being the Schwinger-DeWitt expansion in inverse powers of the field mass.
However, the Schwinger-DeWitt expansion obviously breaks down for massless theories, like our action~\eqref{eq:frame1}, only being able to produce the divergent part of the quantum action.
\par
A useful tool to study heat kernels in the absence of a mass term is the covariant perturbation
theory~\cite{Barvinsky:1987uw,Barvinsky:1990up}, which is based on an asymptotic expansion in terms of
spacetime curvatures, fibre bundle curvatures (gauge field strengths) and potential terms and can be seen as a
resummation of the Schwinger-DeWitt expansion.
In our case, there are no curvatures present, either in spacetime or in field space, and the trace of the heat kernel
can only be expanded as a series in the potential term $P$ as
\be
\Tr K(s)
=
\frac{1}{(4\,\pi\, s)^\omega}
\int\d^4 x\left[1 + s\,P + s^2\, P\,f(-s\Box)\,P + \mathcal O(P^3)\right]
\ ,
\label{eq:trk}
\ee
where
\be
f(u)
=
\frac12\int_0^1\d t\, e^{-t\,(1-t)\,u}
\ .
\ee
Substituting~\eqref{eq:trk} in Eq.~\eqref{eq:proptime} and changing the order of integration,
we find the one-loop contribution to the 1PI action to second order in the potential term $P$
is given by
\be
\tilde\Gamma^{(1)}[\bar\varphi]
=
\frac{q_{\rm B}^2\,\alpha^2}{4\,(4\,\pi)^\omega}
\int\d^4 x\,\frac{J_{\rm B}}{(1 - 3\,\alpha\,\bar\varphi)^{2/3}}
\left[\frac{1}{2-\omega} + 2 - \log\left(\frac{-\Box}{\mu^2}\right)\right]
\frac{J_{\rm B}}{(1 - 3\,\alpha\,\bar\varphi)^{2/3}}
\ ,
\label{eq:quantum}
\ee
for $\omega\to 2$.
Here $\mu$ is a normalisation mass necessary to make the logarithm dimensionless
and whose arbitrariness reflects the renormalisation arbitrariness of the one-loop effective action.
Note that the divergence is proportional to $(1 - 3\,\alpha\,\bar\varphi)^{-4/3}$, whereas the bare action
only contains $(1 - 3\,\alpha\,\bar\varphi)^{4/3}$, which indicates that the theory is non-renormalisable
and only makes sense as an effective field theory.
Since the only mass parameter present in the theory is the Planck mass $\mpl$,
one indeed expects that the theory breaks down at energies of the order of $\mpl$.
\par
Albeit being non-renormalisable, the quantum action can and must be renormalised order by order
in the loop expansion in order to produce sensible results.
This is done by regarding $\tilde S[\bar\varphi,J_{\rm B}]$ as the bare action and adding counter-terms
in an expansion in powers of $\hbar$, to wit
\be
\tilde S[\bar\varphi,J_{\rm B}]
=
\tilde S^{(0)}[\bar\varphi,J_{\rm B}] 
+ \hbar\, \tilde S^{(1)}[\bar\varphi,J_{\rm B}]
+\mathcal O(\hbar^2)
\ .
\ee
The ultraviolet divergences in Eq.~\eqref{eq:quantum} can then be eliminated with the choice
\beq
\tilde S^{(1)}[\bar\varphi,J_{\rm B}]
=
\tilde S^{(1)}_{\rm R}[\bar\varphi,J_{\rm B}]
-\frac{q_{\rm B}^2\,\alpha^2}{4\,(4\pi)^\omega}
\int\d^4x\, \frac{J_{\rm B}}{(1 - 3\,\alpha\,\bar\varphi)^{2/3}}
\left(\frac{1}{2-\omega} + 2 \right)
\frac{J_{\rm B}}{(1 - 3\,\alpha\,\bar\varphi)^{2/3}}
\ ,
\eeq
where $\tilde S^{(1)}_{\rm R}$ denotes the action with renormalised coupling constants.
Note that we included in $\tilde S^{(1)}$ a non-divergent term, which is local and does not contribute to the infrared physics.
Thus we can eliminate it via a finite renormalisation for convenience.
After renormalisation, the infrared equation of motion is finally given by
\be
\Box\bar\varphi
=
q_{\rm B}\, J_{\rm B}\left(1 - 3\,\alpha\,\bar\varphi\right)^{1/3}
+
\frac{q_{\rm B}^2\,\alpha^3\,\hbar}{16\,\pi^2}\,
\frac{J_{\rm B}}
{(1 - 3\,\alpha\,\bar\varphi)^{5/3}}\,
\left.\log\left(-\frac{\Box}{\mu^2}\right)\right|_{R}
\frac{J_{\rm B}}
{(1 - 3\alpha\bar\varphi)^{2/3}}
\ .
\label{eq:effeq}
\ee
The subscript $R$ in the log operator is a reminder that we must impose retarded boundary conditions
by replacing the Feynman Green's function with the retarded one.
This procedure, albeit seemingly ad-hoc, results from the in-in path integral formalism and it is required
in order to obtain a causal evolution for the mean field~\cite{Barvinsky:1987uw,Schwinger:1960qe,Keldysh:1964ud}.
Since the field coordinates in the new frame are Cartesian, we have $\psi^{-1}(\bar\varphi) = \left<\psi^{-1}(\varphi)\right>$.
Therefore, solutions to Eq.~\eqref{eq:effeq} will correspond to corrections to the Newtonian potential in the vacuum state.
\par
We should note that we followed a bottom-up approach to quantum gravity, starting from a modified Newtonian
theory of gravity, by promoting the Newtonian potential to a scalar field and then quantising it.
Nothing in this construction suggests that the resulting theory exhibits any gauge invariance, thus we need not worry
about the usual complications that arise in gauge theories.
However, from a top-down viewpoint, gauge symmetry is required to account for the background independence of gravity.
The reason why gauge symmetry is not realised in bootstrapped gravity is that we phrased the entire approach
in terms of the bootstrapped potential, which is observable~\footnote{For instance,
the potential determines the radial acceleration of a static particle, which is directly observable.} and therefore gauge-invariant.
Should one try to formulate the theory in terms of non-observable quantities,
gauge fixing conditions are expected to become necessary at the quantum level.
A complete analysis of gauge-invariance requires a reconstruction of the full spacetime metric,
which is however left for future investigations.
\par
Since it does not look possible to solve exactly Eq.~\eqref{eq:effeq}, we will expand the solution in powers of the coupling $q_{\rm B}$
as
\be
\bar\varphi
=
q_{\rm B}\,\bar\varphi^{(1)}
+q_{\rm B}^2\, \bar\varphi^{(2)} + \ldots
\ ,
\ee
and solve Eq.~\eqref{eq:effeq} perturbatively up to second order in $q_{\rm B}$.
For the static potential $\bar\varphi=\bar\varphi(\vec x)$ generated by a static source $J_{\rm B}=J_{\rm B}(\vec x)$,
we find
\begin{align}
\label{eq:qb1}
\nabla^2\bar\varphi^{(1)}
&= J_{\rm B}
\ ,
\\
\label{eq:qb2}
\nabla^2\bar\varphi^{(2)}
&=
-\alpha\, J_{\rm B}\,\varphi^{(1)}
+ \frac{\alpha^3\,\hbar}{16\,\pi^2}\,J_{\rm B} \log\left(-\frac{\nabla^2}{\mu^2}\right) J_{\rm B}
\ ,
\end{align}
whose solution is
\begin{align}
\label{eq:sol1}
\bar\varphi^{(1)}(\vec x)
=
&
\int\d^3 x'\, G(\vec x,\vec x')\,J_{\rm B}(\vec x')
\ ,
\\
\bar\varphi^{(2)}(\vec x)
=
&
-\alpha\int\d^3x'\, G(\vec x,\vec x')\,J_{\rm B}(\vec x')\,\bar\varphi^{(1)}(\vec x')
\nonumber
\\
&
+ \frac{\alpha^3\,\hbar}{16\,\pi^2}\int\d^3x'
\int\d^3x''\, G(\vec x,\vec x')\,J_{\rm B}(\vec x')\, L(\vec x',\vec x'')\,J_{\rm B}(\vec x'')
\ ,
\label{eq:sol2}
\end{align}
where
\be
G(\vec x,\vec x') = -\frac{1}{4\,\pi\,|\vec x-\vec x'|}
\ee
is the Green function for the Laplace operator $\nabla^2$.
Moreover, the kernel $L$ of the log operator,
\be
\log\left(-\frac{\nabla^2}{\mu^2}\right)f(x)
=
\int\d^3x'\, L(\vec x,\vec x')\,f(\vec x')
\ ,
\quad
\forall f(x)
\ ,
\ee
is defined as a pseudo-differential operator acting via the Fourier transform,
\be
L(\vec x,\vec x')
=
\int\frac{\d^3q}{(2\,\pi)^3}\, e^{-i\,\vec q\cdot(\vec x-\vec x')}\,
\log\left(\frac{q^2}{\mu^2}\right)
=
-\frac{1}{2\,\pi\, |\vec x-\vec x'|^3}
\ .
\ee
Let us now look at some examples.
\subsection{A point-like source}
\label{sub1}
For the case of a point-like source of mass $M_0$ and current~\footnote{Notice that we are taking into account the
rescalings~\eqref{phiV} and~\eqref{rhoJ}.}
\be
J_{\rm B}(x) = 4\,\pi\,\sqrt{\gn}\,M_0\,\delta(\vec x)
\ ,
\label{1delta}
\ee
we obtain the standard Newtonian potential~\eqref{Vn} with $M=M_0$ from Eq.~\eqref{eq:sol1} by defining $r\equiv |\vec x|$ and
setting $q_{\rm B}=1$,
\be
\bar V_{\rm N}
=
\sqrt{\gn}\,\bar\varphi^{(1)} = -\frac{\gn\,M_0}{r}
\ .
\label{phiN}
\ee
\par
The correction at order $q_{\rm B}^2$ is likewise obtained from Eq.~\eqref{eq:sol2} and reads
\be
\bar\varphi^{(2)}
=
-
\lim_{\epsilon\to 0}
\frac{\alpha\,\gn\,M_0^2}{r\,\epsilon}
\left(
1
-
\frac{\alpha^2\,\hbar}{8\,\pi^2}\,\frac{1}{\epsilon^2}
\right)
\ ,
\label{eq:div}
\ee
which diverges due to the ultra-localized source~\eqref{1delta}.
One instead expects finite results when the Dirac delta is replaced by an extended source (whose radius is greater than the
Planck length).
This is physically expected because strong quantum gravitational effects become important when one probes Planckian
distances and they cannot be accounted for within the realm of effective field theory.
Nonetheless, these divergences can be removed by any regularisation method or, formally, by choosing an integration
contour which does not enclose the spatial origin.
With that in mind, we will drop them out and focus on finite terms, which therefore leads to vanishing corrections
for the Newtonian potential~\eqref{phiN}.
By mapping $\varphi$ back to the original field $\phi$ according to Eq.~\eqref{eq:redef}, one then precisely recovers 
the classical solution $V_{\rm c}$ in Eq.~\eqref{Vc} with $M=M_0$.
\par
We will show in our next example that the above divergences indeed reflect short-distance effects taking place at
Planckian scales by considering two point-like sources.
This way, an extended body can be simulated, with the distance between the sources determining the body's
size.
We will also recover the result that the mass $M$ in the bootstrapped potential is not equal to the proper mass $M_0$
of the source.
\subsection{Two point-like sources}
\label{sub2}
We will next solve Eqs.~\eqref{eq:sol1} and \eqref{eq:sol2} for two massive point-like sources of equal mass $M_0/2$
located at points of coordinate $\vec x_1$ and $\vec x_2$ and contributing to the total source as
\be
J_{\rm B}(x) = 2\,\pi\,\sqrt{\gn}\,M_0\left[\delta(\vec x-\vec x_1) + \delta(\vec x-\vec x_2)\right]
\ .
\ee
Eq.~\eqref{eq:sol1} gives again the classical Newtonian potential
\be
\sqrt{\gn}\,\bar\varphi^{(1)}
=
-\frac{\gn\,M_0}{2}\left(\frac{1}{|\vec x-\vec x_1|} + \frac{1}{|\vec x-\vec x_2|}\right)
\ .
\ee
If we assume that $|\vec x_1-\vec x_2|\equiv R\ll |\vec x-\vec x_1|\simeq |\vec x-\vec x_2|\equiv r$, we can
in fact write
\be
\sqrt{\gn}\,\bar\varphi^{(1)}
\simeq
-\frac{\gn\,M_0}{r}
\ ,
\ee
which reproduces the result in the previous example to first order in $q_{\rm B}$.
\par
For the calculation of $\bar\varphi^{(2)}$, we must point out that the product of quantities evaluated at the same point
$\vec x_i$, such as $\delta(\vec x-\vec x_i)\,\bar\varphi^{(1)}(\vec x_i)$, is ill-defined.
This is analogous to the situation of the previous example and only reflects the existence of short-distance effects
beyond the grasp of the effective field theory.
We will therefore drop single-point quantities and focus on the cross terms of quantities evaluated at $\vec x_1$ and $\vec x_2$.
Eq.~\eqref{eq:sol2} then gives
\be
\bar\varphi^{(2)} 
=
-\frac{\alpha\,\gn\, M_0^2}{2\,|\vec x_1-\vec x_2|}
\left(
1
-
\frac{\alpha^2\,\hbar}{8\,\pi^2\,|\vec x_1-\vec x_2|^2}
\right)
\left(\frac{1}{|\vec x-\vec x_1|} + \frac{1}{|\vec x-\vec x_2|}\right)
\ .
\ee
Finally, the total solution to order $q_{\rm B}^2$ reads
\begin{align}
\sqrt{\gn}\,\bar\varphi
\simeq
&
-\frac{q_{\rm B}\,\gn\,M_0}{2}
\left[1 
+\frac{q_{\rm B}\,q_\phi\,\gn\, M_0}{|\vec x_1-\vec x_2|}
\left(1
- \frac{q_\phi^2\,\lp\,\hbar}{2\pi^2\,\mpl\, |\vec x_1-\vec x_2|^2}\right)
\right]
\left(\frac{1}{|\vec x-\vec x_1|} + \frac{1}{|\vec x-\vec x_2|}\right)
\nonumber
\\
\simeq
&
-\frac{\gn\,M_0}{r}
\left[1 
+\frac{q_\phi\,\gn\, M_0}{|\vec x_1-\vec x_2|}
\left(1
- \frac{q_\phi^2\,\lp^2}{2\pi^2\, |\vec x_1-\vec x_2|^2}\right)
\right]
\ ,
\label{bvphi}
\end{align}
in which we used Eq.~\eqref{alpha} to display the original coupling $q_\phi$ and again set $q_{\rm B}=1$ at the end.
Note that one recovers the classical Newtonian potential either for $q_\phi=\alpha = 0$ or, more formally, when the sources
are far away from each other, $|\vec x_1-\vec x_2|\to\infty$.
At first order in $q_\phi\sim \alpha$ we have the classical contribution to the non-linearity introduced in the action~\eqref{eq:s0},
whereas at third order we find the correction due to (one-loop) quantum gravity.
We are also able to interpret the divergences in Eq.~\eqref{eq:div} for a single point-like particle as indeed originated
from the limit $\vec x_1\to \vec x_2$ in Eq.~\eqref{bvphi}.
Although we still had to deal with the single-point divergences mentioned above, they will presumably yield finite
results once a smooth matter source comprising of all macroscopical sources, is used as opposed to a system
of point-like sources.
\par
Note that the above expression~\eqref{bvphi} more accurately reproduces the classical solution $V_{\rm c}$ in Eq.~\eqref{Vc} outside
an extended source of small compactness $\gn\,M_0\ll R$.
In fact, we can consider the distance between the two point sources as the size $R$ of an extended source,
that is $R\simeq |\vec x_1-\vec x_2|$, and introduce the modified mass 
\begin{align}
\tilde M
&
\equiv
M_0
\left[1 + q_\phi\,\frac{\gn\, M_0}{R}
\left(1 - \frac{q_\phi^2\,\lp^2}{2\pi^2\,R^2}\right) 
\right]
\nonumber
\\
&
\simeq
M
\left(
1- \frac{q_\phi^2\,\lp^2}{2\pi^2\,R^2}
\right)
\ ,
\label{M-M0}
\end{align}
where the relation between $M$ and $M_0$ is the same as the one found in Ref.~\cite{Casadio:2018qeh}
for a uniform star of small compactness, modulo a numerical coefficient of order one.
We then obtain
\be
\sqrt{\gn}\,\bar\varphi
\simeq
-\frac{\gn\,\tilde M}{r}
\ .
\label{bvphin}
\ee
By transforming back to the original frame $\sqrt{\gn}\,\phi= V$, we then recover the classical bootstrapped potential~\eqref{Vc}
in the vacuum with the mass $M$ now further corrected by a one-loop contribution.~\footnote{We note in passing that this contribution
is consistently of the same order $\lp^2$ as the corrections found in Ref.~\cite{Calmet:2019eof} for the metric generated by a star.}
We remark that the expression for the mass~\eqref{M-M0} only holds for small compactness of the source (that is for
$\gn\,M_0/R\ll 1$) and that $M>M_0$ is precisely a consequence of the nonlinearity included in the bootstrapped 
dynamics, as discussed in Refs.~\cite{Casadio:2018qeh,Casadio:2019cux}.
\par
Finally, we must emphasize that the above solution is non-perturbative in $q_\phi$ and should reproduce all effects
due to the non-linear self-coupling of gravity.
On the other hand, the coupling to matter was handled perturbatively.
Treating both $q_\phi$ and $q_{\rm B}$ non-perturbatively is utterly difficult, but one can study non-perturbative effects
due to the gravitational self-coupling at the expense of dealing perturbatively with respect to the coupling to matter.
\section{Conclusions}
\label{sec:conc}
\setcounter{equation}{0}
In this paper, we have considered non-linear derivative self-interactions of the Newtonian potential
by allowing the first few post-Newtonian terms to take arbitrary values.
Such a theory has been called bootstrapped Newtonian gravity.
We calculated one-loop quantum corrections to the bootstrapped potential by first promoting
the non-relativistic potential to a Lorentz covariant form that allows the application of quantum field theory
techniques in intermediate steps.
These intermediate calculations are obviously supposed to serve only as a guideline for the quantisation of the
complete theory of gravity ({\em e.g.}~light bending requires a non-scalar gravitational field), thus we must take
the non-relativistic limit in the end, which is all we need for our purposes.
\par
We showed that the bootstrapped Newtonian potential is described by a non-local equation of motion in the infrared,
which is typical of massless theories.
We solved it for a point-like source and a system of two point-like sources.
The latter can be thought of as a rough approximation of an extended source.
Our results recover Newtonian physics in the limit where the sources are far apart and for vanishing derivative
interactions $\alpha\to 0$, as one would expect.
The analysis of more realistic situations, such as a smooth extended source,
proves much more challenging already at the classical level~\cite{Casadio:2019cux}
and will be left for future investigations.
In any case, the effective equations of motion together with the resulting quantum bootstrapped potential
permits a better understanding of quantum processes taking place at non-perturbative settings which are
important for strong field applications.
\section*{Acknowledgments}
R.C.~and I.K.~are partially supported by the INFN grant FLAG.
The work of R.C.~has also been carried out in the framework of activities of the National Group of Mathematical Physics
(GNFM, INdAM) and COST action Cantata.
%
% 
%%%%%%%%%%%%%%%%%%%%%%%%%%%%%%%%%%%%%%%%%%%%%%%%%%%%%%%%%%%%%%%%%
%%%
%%%                     BIBLIOGRAPHY
%%%
%%%%%%%%%%%%%%%%%%%%%%%%%%%%%%%%%%%%%%%%%%%%%%%%%%%%%%%%%%%%%%%%%
%
%

%
\end{document}